\documentclass[final,12pt]{elsarticle}

\usepackage{graphics} % for pdf, bitmapped graphics files
\usepackage{color,soul}
\usepackage{xcolor}

\usepackage{epsfig} % for postscript graphics files
\usepackage{times} % assumes new font selection scheme installed
\usepackage{amsmath} % assumes amsmath package installed
\usepackage{amssymb}  % assumes amsmath package installed

\usepackage{xcolor}

\begin{document}
	
	\begin{frontmatter}
		
		
		
		\title{Grasping force estimation using state-space model and Kalman Filter}
		
		\author[label1]{DUTRA, B. G.}
		\ead{brunodutra@ufpa.br}
		\author[label2]{SILVEIRA, A. S.}
		\ead{asilveira@ufpa.br}
		\author[label3]{PEREIRA, A.}
		\ead{apereira@ufpa.br}
		
		\address[label1,Label2,label3]{{Federal University of Par\'a}, {Bel\'em}, {66075-110}, {Brazil}}

		\begin{abstract}
			The grip force required to handle an object depends on the object's mass and the friction coefficient of its surface. The control of grip force in myoelectric prosthesis is crucial for handling objects adequately. In the current paper we propose a new method for improving the proportional and continuous grasping force estimation to improve control systems for myoelectric prosthesis based on surface electromyography (sEMG) recordings. For this purpose, we develop an approach based on multivariable system identification in the state-space (SS) and continuous force estimation with Kalman Filter (KF). The sEMG recordings of ten healthy individuals performing a grip task were used as data set for model identification. The root mean square (RMS), the mean absolute value (MAV), and the waveform length (WL) extracted from the sEMG signals were used at the model's input while the measured grasping force was the output. The performance of the method was evaluated with the normalized root-mean-squared-error (NRMSE) and the square of the Pearson’s correlation coefficient ($ R^2 $). We found the $R^2$ and NRMSE values were 0.92$ \pm $ 0.0319 and 0.723$ \pm $ 0.0563, respectively. The performance of the proposed technique was superior to the results obtained with other regression models, such as the recurrent nonlinear autoregressive exogenous (NARX)-based neural network, the multi-layer perceptron (MLP) network and the  linear discriminant analysis (LDA) with a quadratic polynomial fitting (QPF). The results confirm that the method is adequate for real-time applications with myoelectric hand prostheses.
			
		\end{abstract}
		
		\begin{keyword}
			%% keywords here, in the form: keyword
			Electromyography
			\sep Grasping force
			\sep Kalman Filter
			\sep State-space identification
			
			
		\end{keyword}
		
	\end{frontmatter}
	
	% \linenumbers
	
	\section{Introduction}
	\label{Introduction}
	Surface electromyography (sEMG) is a noninvasive recording method which can be used for the implementation of prostheses controlled by muscles of the residual limb \cite{scheme2011electromyogram}. More recently, research on this subject has focused on improving prosthesis control and make it more intuitive to patients \cite{scheme2011electromyogram,marasco2018illusory,igual2019myoelectric}. Ideally, prosthetic systems aimed for upper limb replacement should allow subjects to precisely control their grasping force when manipulating objects. Thus, it is first necessary to estimate the user's grasping force from electrophysiological recordings from units in the descending motor pathways and use it to control the prosthesis.
	
	The force required to grasp an object is proportional to the energy of sEMG signals in a group of forearm muscles \cite{hudgins1993new,phinyomark2012feature}. However, crosstalking interference makes it difficult to directly relate the signals recorded from a specific muscle with grasping force \cite{winter1994crosstalk}. Also, the structural and physiological differences among users is a complicating factor regarding the accurate estimation of grasping force.
	
	Several studies focused on improving estimations of grasping force from sEMG recordings from arm muscles. In those attempts, researchers have relied on mathematical methods based on biomechanics \cite{hill1938heat,clancy1997relating,menegaldo2012exploring}, polynomial equations \cite{clarke1979self,kamavuako2012estimation}, machine learning techniques \cite{castellini2009multi,choi2010real, zhuojun2015semg, zhang2017pattern} and system identification \cite{kumar2010adaptive,ohno2017motion}.
	For instance, Castellini and coworkers (2009) obtained $90\%$ accuracy in force estimation with the use of a support vector machine (SVM) based on inputs from six sEMG channels  \cite{castellini2009fine}. Potluri and coworkers (2015) used an optimized linear model fusion algorithm for continuous force estimation from 3 sEMG channels and obtained an average correlation of $85,6\%$ with the original force value to provide closed-loop feedback control for a robotic finger  \cite{potluri2015real}. 
	
	In a recent study, Ohno and coworkers (2017) utilized the nonlinear autoregressive exogenous (NARX) model to estimate the grasping force and the angle of the wrist from sEMG signals with high accuracy even under different sampling rates \cite{ohno2017motion}. Li and coworkers (2018) proposed a system for the control of grasping force based on dimensionality reduction with principal component analysis (PCA) and estimation of sEMG energy with a neural network (NN). The authors obtained an average accuracy of $95\%$ for eight levels of grip strength \cite{li2018pca}. In another study, Wang and coworkers (2019) used linear discriminant analyses (LDA) with a quadratic polynomial model to reduce the dimensionality of multi-channel sEMG recordings and showed that it provided continuous estimation of the grasping force with a goodness-of-fit of $82.05\%$ for real-time implementation \cite{wang2019recognition}. Ma and coworkers (2020) compared the performance of a gene expression programming (GEP) algorithm with a multi-layer perceptron (MLP) for estimating grasping force from inputs from 16 sEMG channel recordings and found the GEP's performed best \cite{ma2020grasping}.
	
	In the present work, we present a new method for the real-time estimation of grasping force from sEMG measurements from upper limb muscles aimed at the real-time control of a hand prosthesis. Thereby, the main contribution of this paper is the proposal of a new model for proportional and continuous grasping force estimation to improve myoelectric control systems. The proposed model is based on multivariable system identification in the state-space (SS) with a Kalman Filter (KF) tuned to exploit optimal results in the minimum variance sense and improve both system identification and the the acquisition of dynamic information based on process uncertainties
	\cite{astrom1971introduction,silveira2020design}.
	We propose a MISO (Multiple-Inputs Single-Output) model for grasping-force estimation that uses the energy characteristics of the sEMG signals as input. To demonstrate the validity of the method, we compare its performance to other methods proposed in the literature, such as the nonlinear MLP model, the recurrent NARX model based on NN, and the LDA method with a quadratic polynomial fitting (QPF).
	
	Our results show that:
	\begin{itemize}
		\item The proposed model is stable and accurate, with low volatility during the real-time estimation of the grasping force based on sEMG inputs.
		\item The proposed model has the highest Normalized-Root-Mean-Squared-Error (NRMSE) and a higher 
		square of Pearson’s correlation coefficient ($ R^2 $) than the MLP, the NARX model, and the LDA/QPF model.
		\item  Both the computational cost and the memory requirements of the model are relatively low and provide a good option to be implemented with low-end microcontrollers for real-time applications with myoelectric prosthetic systems.
	\end{itemize}
	
	%Beyond this introductory part, this work is organized as follows:
	%Section \ref{sinals} presents the procedure for data acquisitions, experimental process, and feature extraction. Section \ref{force model} presents the proposed method of identifying the SS MISO linear grasping force model, the Kalman Filter tuning, and the use of the MLP and NARX model to estimate the grasping force. Section \ref{results} shows the experimental results of grasping force estimation and discusses it. Finally, the paper ends with a conclusion and future works in Section \ref{conclusion}.

	\section{Materials and Methods}
	\label{sinals}
	
	\subsection{sEMG and force data acquisition}
	\label{data acquisition}
\begin{figure}[thpb]
		\centering
		\includegraphics[scale=0.6]{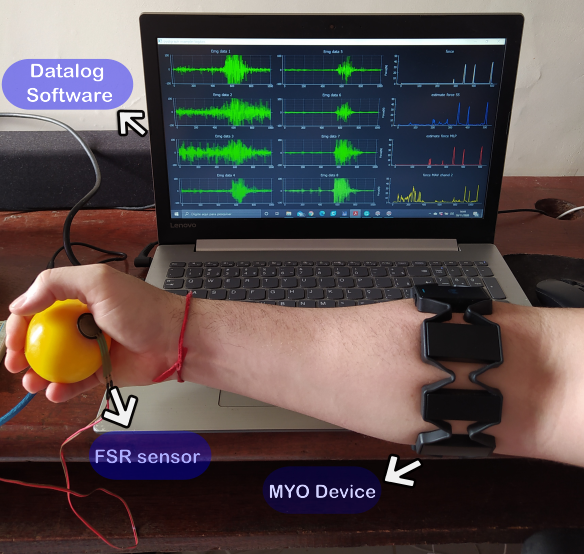}
		\caption{Experimental Setup.}
		\label{fig:setup}
	\end{figure}

	The system used for datalog and real-time visualization is shown in Fig. \ref{fig:setup}.
	The MYO armband (Thalmic Labs, Kitchener-Waterloo, Canada) was used to record sEMG signals from the upper limb. The MYO has eight sEMG input channels, wireless communication via Bluetooth protocol, sampling rate of 200 Hz and 8-bit resolution. Also, a force-sensitive-resistor (FSR) (Model 402, Interlink Electronics, Inc.) was used for force measurement with sensitivity from 1N to 50N. The FSR was inserted in an anti-stress ball and its inputs were fed into an Arduino-based data-acquisition (Daq) device, with a 10-bit analog-to-digital resolution. The datalog devices communicate with a Windows PC running Python for real-time visualization and data storage.
	
	\subsection{Experimental process}
	\label{protocol}
\begin{figure}[thpb]
		\centering
		\includegraphics[scale=0.45]{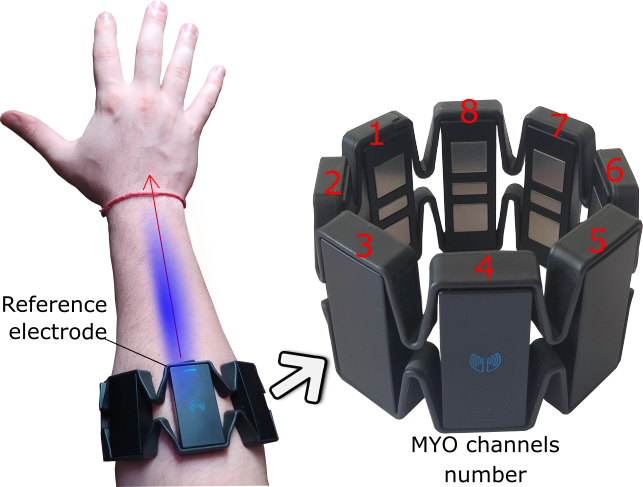}
		\caption{Reference electrode and channels numbers of MYO.}
		\label{reference_channels}
	\end{figure}

	The experimental procedures were approved by the local ethics committee of the Federal University of Pará (82131517.1.0000.0018). A total of 10 healthy voluntaries (four females and six males, with 24 $\pm$ 4 years old) were selected for the grasping force estimation study and signed an informed consent. The MYO device was placed in the same forearm location for each subject. Channel 4 was used as reference and was positioned in the extensor digitorum muscle, as shown in Fig. \ref{reference_channels}. The location of superficial electrodes over forearm muscles is detailed in Table \ref{ref_muscle}.
	
\begin{figure}[thpb]
		\centering
		\includegraphics[scale=0.5]{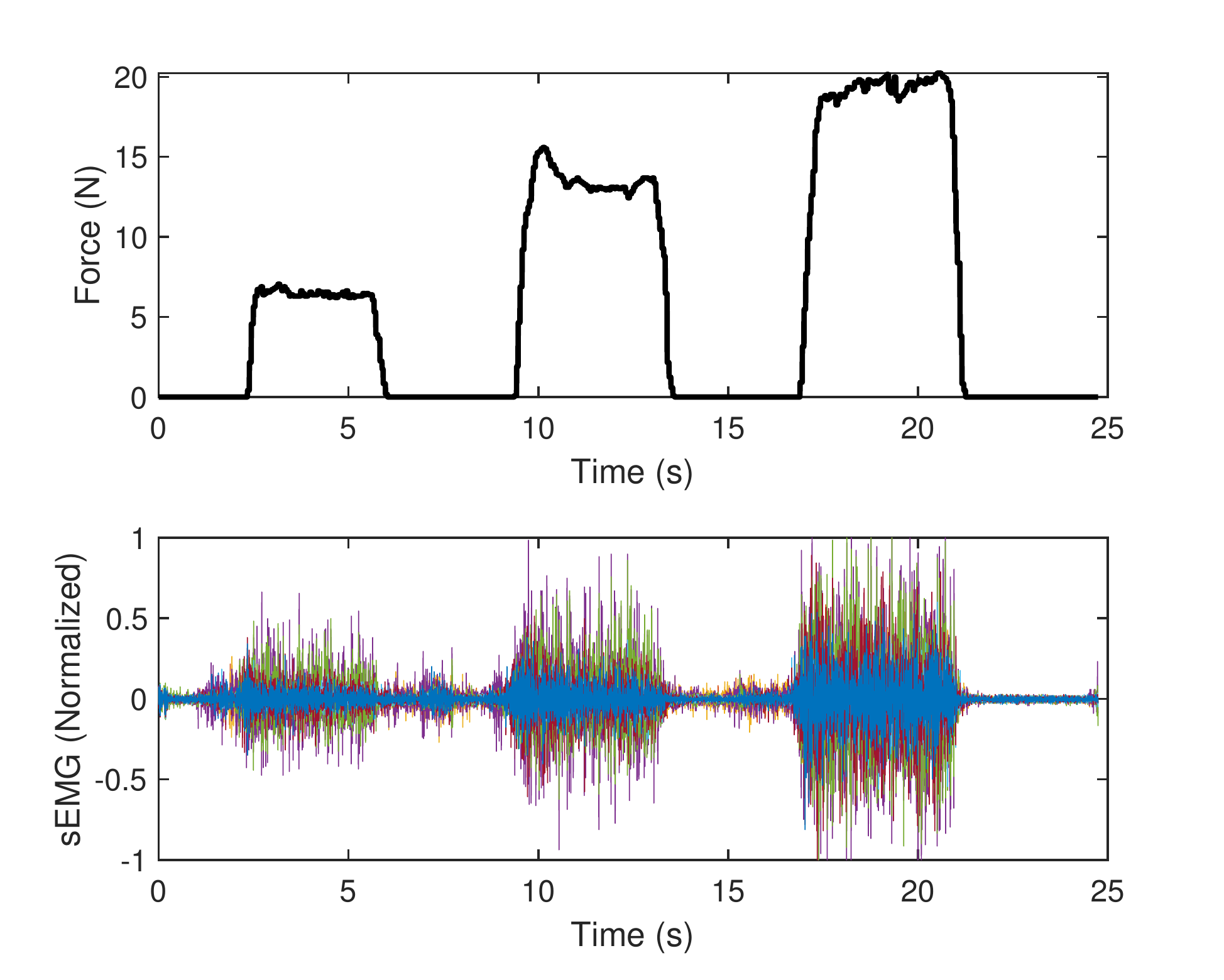}
		\caption{Data log of MVC grasping force and sEMG signals from volunteer 10.}
		\label{datalogs}
	\end{figure}

	At the beginning of the experiment, each participant received the following command: "Position the ball with the sensor facing the palm, then press the ball three times. The first time squeeze it with a small force, the second time with a medium  force, and the third time with the highest force. Pay attention to the commands to start and end the movement". The sEMG and the force exerted by each volunteer were simultaneously recorded within the 0-100 \% range of the maximum voluntary contraction (MVC), as shown in Fig. \ref{datalogs}. 
	
	\begin{table}[htb]
		\centering
		\begin{tabular}{ll}
			\hline
			\multicolumn{1}{|c|}{Localization} & \multicolumn{1}{r|}{Muscle name} \\ \hline
			(1)                                                       & Pronator Teres                       \\
			(2)                                                       & Brachioradialis                       \\
			(3)                                                       & Extensor Carpi Radialis             \\
			(4)                                                       & Extensor Digitorum Cummunis         \\
			(5)                                                       & Exstensor Carpi Ulnaris             \\
			(6)                                                       & Flexor Carpi Ulnaris               \\
			(7)                                                       & Palmaris Longus                         \\
			(8)                                                       & Flexor Carpi Radiallis
		\end{tabular}
		\caption{Distribution of the electrodes on the forearm muscles.}
		\label{ref_muscle}
	\end{table}
	
	\subsection{Feature selections}
	\label{features}
	Some sEMG features can be correlated with the magnitude of the grasping force \cite{hakonen2015current,phinyomark2012feature}. In this paper, we chose three sEMG features to assess the accuracy of grasping force prediction in the time-domain: The mean absolute value (MAV), the root mean square (RMS), and the waveform length (WL).
	
	The MAV provides an energetic estimate of sEMG by averaging the sum of all $ x (k) $ samples within a range of $ N $ samples:
	\begin{equation} \label{Equation_3_2}
		MAV = \frac{1}{N}\sum\limits_{k = 1}^N {|x(k)|} 
	\end{equation}
	where $ x (k) $ is the sEMG signal at time $ k $ and $ N $ is the number of samples obtained in the window interval.
	
	The RMS is also an energetic estimate of the sEMG. It is a feature related to the magnitude of the sEMG and is calculated by the root-mean-square of $ x (k) $ samples:
	\begin{equation} \label{Equation_3_3}
		RMS = \sqrt {\frac{1}{N}\sum\limits_{k = 1}^N {{x^2}(k)} } 
	\end{equation}
	
	WL gives information about the complexity of the signal in a window by summing the numerical derivative of the sample window  \cite{hudgins1993new}. This is represented by the cumulative length of the waveform over the time segment defined as:
	\begin{equation} \label{Equation_3_4}
		WL = \sum\limits_{k = 1}^N {|x(k) - x(k - 1)|} 
	\end{equation}
	
	We used a pre-processing window to extract the sEMG features. The window length directly affects the performance and the stability of the grasping force estimation \cite{englehart2003robust}. However, for real-time applications, the window length update should be less than 300 ms to meet delay requirements \cite{hudgins1993new}. Therefore, we used an overlapping window of 400 ms with an increment of 125 ms.
	
	\section{Grasping force model}
	\label{force model}
	
	\subsection{State-Space identification }
	\label{SS identi}
	We used the recursive least squares (RLS) algorithm to estimate the grasping force from sEMG signals  with the following state-space model:
	\begin{equation} \label{Equation2}
		x(k) = Ax(k - 1) + Bu(k - 1) + \Gamma w(k - 1)% 
	\end{equation}
	\begin{equation} \label{Equation3}
		y(k) = Cx(k) + v(k)
	\end{equation}
	where $ x(k) $ is the state vector, $ u (k) $ is the input of the system, $ y (k) $ is the output of the system and $ k $ is the discrete-time domain variable. $ A $ is the state matrix, $ B $ is the input weighting matrix, $\Gamma $ is the state matrix related to the Gaussian noise, $ w(k)$ is the process noise, $C$ is the matrix that associates the states with the measured output of the process, and $ v(k) $ is the measurement noise.
	
	The input $ u(k) $ and output $ y(k) $ are the variables used to identify the model and the state variable $ x(k) $ can be estimated using the subspace method, given that all state variables are  measurable \cite{van2012subspace,ljung1996subspace}. Assuming that the measured output $y$ correspond to the state variable $x_1$, the estimation of the next state, $x_2$, can be calculated by the backward difference of $x_1$, and so on. For a higher-order model the other state variables can be calculated by the  approximation of the discrete derivative of the previous state variable as a function of time, as below:  
	
	\begin{equation} \label{Equation4_3}
		\left[ {\begin{array}{*{20}{c}}
				{{x_1}(k)}\\
				{{x_2}(k)}\\
				\vdots \\
				{{x_n}(k)}
		\end{array}} \right] = \left[ {\begin{array}{*{20}{c}}
				{y(k)}\\
				{\frac{{{x_1}(k) - {x_1}(k - 1)}}{{T_s}}}\\
				\vdots \\
				{\frac{{{x_{n-1}}(k) - {x_{n-1}}(k - 1)}}{{T_s}}} 
		\end{array}} \right]
	\end{equation} 
	
	Considering a generic SS model with order $ n $, $ i $ inputs and $ j $ outputs
	\begin{equation} \label{Equation4_4}
		\begin{array}{l}
			\underbrace {\left[ {\begin{array}{*{20}{c}}
						{{x_1}(k)}\\
						\vdots \\
						{{x_n}(k)}
				\end{array}} \right]}_{x(k)} = \underbrace {\left[ {\begin{array}{*{20}{c}}
						{{a_{11}}}& \cdots &{{a_{1n}}}\\
						\vdots & \ddots & \vdots \\
						{{a_{n1}}}& \cdots &{{a_{nn}}}
				\end{array}} \right]}_A\underbrace {\left[ {\begin{array}{*{20}{c}}
						{{x_1}(k - 1)}\\
						\vdots \\
						{{x_n}(k - 1)}
				\end{array}} \right]}_{x(k - 1)} + \underbrace {\left[ {\begin{array}{*{20}{c}}
						{{b_{11}}}& \cdots &{{b_{1i}}}\\
						\vdots & \ddots & \vdots \\
						{{b_{n1}}}& \cdots &{{b_{ni}}}
				\end{array}} \right]}_B\underbrace {\left[ {\begin{array}{*{20}{c}}
						{{u_{1}}(k - 1)}\\
						\vdots \\
						{{u_{i}}(k - 1)}
				\end{array}} \right]}_{u(k - 1)}\\
			\,\,\,\,\,\,\,\,\,\,\,\,\,\,\,\,\,\,\,\,\,\,\,\,\,\,\,\,\,\,\,\,\,\,\,\,\,\,\,\,\,\,\,\,\,\,\,\,\,\, + \underbrace {\left[ {\begin{array}{*{20}{c}}
						{{\gamma _{11}}}& \cdots &{{\gamma _{1n}}}\\
						\vdots & \ddots & \vdots \\
						{{\gamma _{n1}}}& \cdots &{{\gamma _{nn}}}
				\end{array}} \right]}_\Gamma \underbrace {\left[ {\begin{array}{*{20}{c}}
						{{w_1}(k - 1)}\\
						\vdots \\
						{{w_n}(k - 1)}
				\end{array}} \right]}_{w(k - 1)}
		\end{array}
	\end{equation}
	\begin{equation} \label{Equation4_5}
		\underbrace {\left[ {\begin{array}{*{20}{c}}
					{{y_1}(k)} \\ 
					\vdots  \\ 
					{{y_j}(k)} 
			\end{array}} \right]}_{y(k)} = \underbrace {\left[ {\begin{array}{*{20}{c}}
					{{I_{j \times j}}}&{{0_{j \times (n - j)}}} 
			\end{array}} \right]}_C\underbrace {\left[ {\begin{array}{*{20}{c}}
					{{x_1}(k)} \\ 
					\vdots  \\ 
					{{x_n}(k)} 
			\end{array}} \right]}_{x(k)} + \underbrace {\left[ {\begin{array}{*{20}{c}}
					{{v_1}(k)} \\ 
					\vdots  \\ 
					{{v_j}(k)} 
			\end{array}} \right]}_{v(k)}
	\end{equation}
	and knowing that $ y(k) $, $ x(k) $ and $ u(k) $ are available, the estimated parameters matrix can be defined as:
	
	\begin{equation}\label{eq_params_vector}
		\hat \theta  = {\left[ {\underbrace {\begin{array}{*{20}{c}}
						{{a_{11}}}& \cdots &{{a_{1n}}}\\
						\vdots & \ddots & \vdots \\
						{{a_{n1}}}& \cdots &{{a_{nn}}}
				\end{array}}_A\underbrace {\begin{array}{*{20}{c}}
						{{b_{11}}}& \cdots &{{b_{1i}}}\\
						\vdots & \ddots & \vdots \\
						{{b_{n1}}}& \cdots &{{b_{ni}}}
				\end{array}}_B\underbrace {\begin{array}{*{20}{c}}
						{{\gamma _{11}}}& \cdots &{{\gamma _{1n}}}\\
						\vdots & \ddots & \vdots \\
						{{\gamma _{n1}}}& \cdots &{{\gamma _{nn}}}
				\end{array}}_\Gamma } \right]^T}
	\end{equation}
	
	The future observation vectors are organized by the following  vector of regressors:
	\begin{equation}\label{Equation4_7}
		{\phi ^T}(k) = [x^T{(k - 1)}\,u^T(k - 1)\,\,w^T{(k - 1)}]
	\end{equation}
	where the vector $ w^T{(k)} $, related to the process noise, is calculated for every instant $k$ by the RLS algorithm estimation error and has null values as initial conditions.
	
	Equation (\ref{Equation4_7}) allows the recursive identification of $ {\hat \theta} $. Then, in an interactive way with the data, the estimated states are calculated by
	\begin{equation} \label{Equation4_8}
		\hat x(k) = {\varphi ^T}(k)\hat \theta (k - 1).
	\end{equation}
	Next, the estimated model output is calculated by
	\begin{equation} \label{Equationx}
		\hat y(k) = C\hat x(k),
	\end{equation}
	and then the prediction error is obtained as follows:
	\begin{equation} \label{Equation4_9}
		e(k) = x(k) - \hat x(k).
	\end{equation}
	The estimator gain is given by
	\begin{equation} \label{Equation4_10}
		K(k) = \frac{{P(k - 1)\varphi (k)}}{{1  + {\varphi ^T}(k)P(k - 1)\varphi (k)}},
	\end{equation}
	which allows the matrices of parameters to be updated by
	\begin{equation} \label{Equation4_11}
		\hat \theta (k) = \hat \theta (k - 1) + K(k)e(k).
	\end{equation}
	Based on the estimation uncertainties, the process noise is obtained by
	\begin{equation} \label{Equation4_12_2}
		w(k) = x(k) - {\varphi ^T}(k)\hat \theta (k).
	\end{equation}
	Similarly, the measurement noise is:
	\begin{equation} \label{Equation4_12}
		v(k) = y(k) - \hat y(k).
	\end{equation}
	
	The covariance matrix of the estimator is updated by
	\begin{equation} \label{Equation4_13}
		P(k) = \left[I - K(k){\varphi ^T}(k)\right]P(k - 1),
	\end{equation}
	and Eqs. (\ref{Equation4_8}) to (\ref{Equation4_13}) are solved recurrently for each increase in time instant $k$ until the end of the data log. $ \hat \theta (0) $ can be initialized experimentally with low values and was adjusted to 0.3 for its elements. $P(0)$ was initialized by $ P(0) = m{I_{(2n + i)\, \times \,(2n + i)}} $, where $ m $ is a high value $ (m \approx {10^3}$ or even higher$ )$, assuming that the power of the initial uncertainty is sufficiently larger than the steady-state uncertainty obtained after the RLS algorithm has converged to the optimal parameters.
	
\begin{figure}
		\centering
		\includegraphics[scale=0.21]{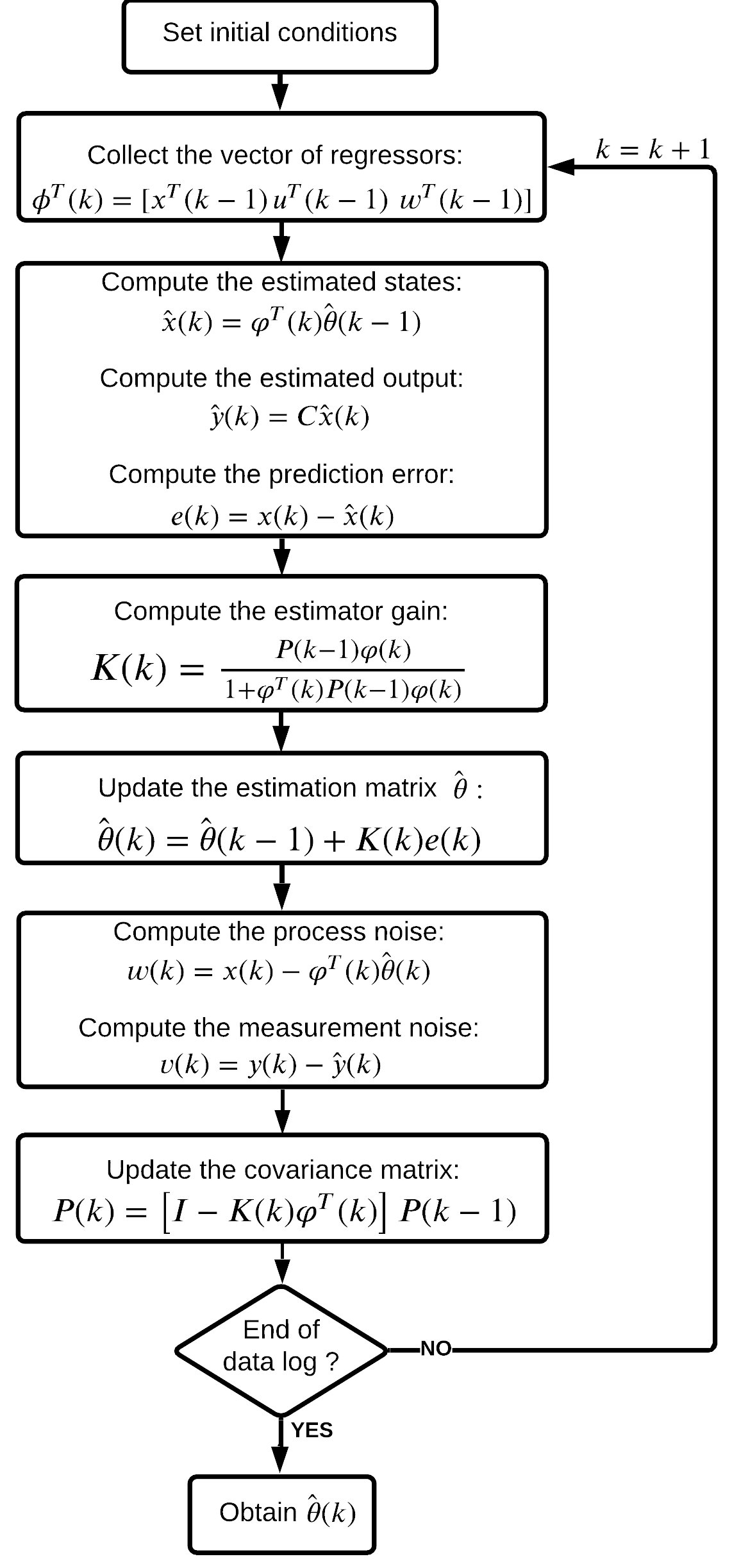}
		\caption{Flowchart of RLS algorithm to obtain the matrix $ \hat \theta$.}
		\label{flowchart}
	\end{figure}

	The steps of the RLS algorithm until the convergence of parameters and update of the matrix $ \hat \theta$, are presented as a flowchart in Figure \ref{flowchart}.

	%Finally, after the parameters' convergence, the matrix $ \hat \theta$ will have the values of the estimated model.
	
	\subsection{Kalman Filter}
	\label{Kalman}
	The Kalman Filter can estimate the full state variables, provide data fusion, filtering, and minimum variance approximation \cite{astrom1971introduction,lewis2017optimal,ljung1996subspace}. The KF is based on the SS model presented in (\ref{Equation2}) and (\ref{Equation3}).
	
	As a dynamical system, the KF can be related to the following state estimator structure \cite{lewis2017optimal}:
	\begin{equation} \label{Equation4_18}
		\hat x(k) = \left( {A - LC} \right){\hat x}(k - 1) + Bu(k - 1) + L y(k-1)
	\end{equation}
	\begin{equation} \label{Equation4_19}
		{y_{KF}}(k) = C\hat x(k)
	\end{equation}
	where $\hat x(k)$ is the estimated state vector and $ {y_{KF}}(k) $ is the KF estimated output.
	
	The optimal gain
	\begin{equation} \label{Equation 16}
		L = AP_{KF}{C^T}{\left( {CP_{KF}{C^T} + R_{KF}} \right)^{ - 1}},
	\end{equation}
	can be obtained offline, for the non-adaptive case, by iterating the estimator Riccati difference equation \cite{lewis2017optimal,goodwin1984k} and calculating the covariance matrix
	\begin{equation} \label{Equation 17}
		\begin{array}{l}
			P_{KF}(k + 1) = AP_{KF}(k){A^T}
			- AP_{KF}(k){C^T}\\
			{\left( {CP_{KF}(k){C^T} + {R_{KF}}} \right)^{ - 1}}CP_{KF}(k){A^T} + {Q_{KF}},
		\end{array}
	\end{equation}
	that minimizes the estimation error given by $e_{est}=x(k)-\hat x(k) $,
	based on the result of $ P_{KF}: = P_{KF}(k \to \infty ) $, starting with a high magnitude $P_{KF}(0)$.

	With KF's optimal estimation and minimum variance approximation, it is possible to mitigate identification errors and to filter out noise components so that the state variables have the best possible correction, while reducing the mean squared error. For the minimum variance case, the KF weighting matrices $Q_{KF}$ and $R_{KF}$ are tuned according to the covariance of the process noise and the covariance of the measurement noise \cite{lewis2017optimal}, respectively:
	
	\begin{equation} \label{Equation4_20}
		{Q_{FK}} = \Gamma Q{\Gamma ^T},
	\end{equation}
	\begin{equation} \label{Equation4_21}
		R_{FK} = R,
	\end{equation}
	where $Q = diag(\begin{array}{*{20}{c}}
		{\sigma^2_{w_1}}& \cdots &{\sigma^2_{w_n}} 
	\end{array})$, $R = diag(\begin{array}{*{20}{c}}
		{\sigma^2_{v_1}}& \cdots &\sigma^2_{v_{j}} 
	\end{array})$. The matrix describing the Gaussian noise input, $\Gamma$, is estimated together with matrices $A$ and $B$, since these matrices are included in the matrix $ \hat \theta $ shown in (\ref{eq_params_vector}). The estimation of $\Gamma$ occurs iteratively, with the RLS algorithm, through estimation of the process noise $w(k)$, as shown in (\ref{Equation4_12_2}). 
	
	\subsection{Force model identification}
	\label{Model_identificaiton}
	To establish the state-space model, we used the sEMG features, MAV, RMS, and WL as inputs and the measured grasping force as the model's output, all normalized from 0 to 1. Thus, for the grasping force model's identification, the system's order and the number of inputs are defined based on the number of sEMG channels multiplied by the number of features used per channel.
	
\begin{figure}[thpb]
		\centering
		\includegraphics[scale=0.28]{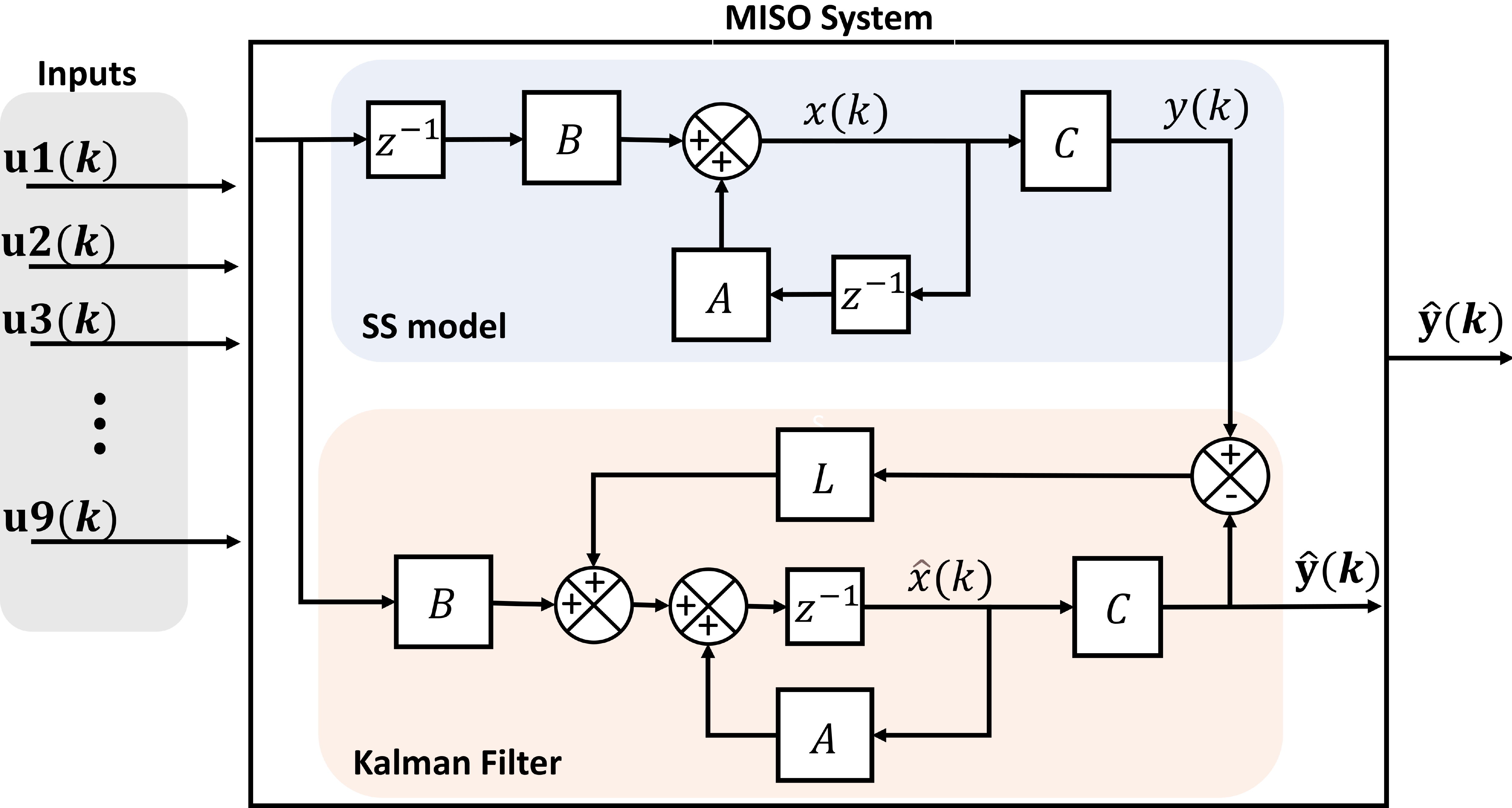}
		\caption{Structure of the state-space MISO grasping force model with Kalman Filter.}
		\label{miso model}
	\end{figure}

	%Considering the relationship between sEMG features and the grasping force, there is the possibility of using all eight channels of the MYO device or its combinations.
	%The main muscles responsible for Grip Strength are the Superficial Flexor, Flexor Digitorum Profundus, and the Flexor Pollicis Longus. However, since these muscles are located superficially in the forearm \cite{kendall1995musculos},
	We selected the sEMG channels corresponding to the superficial muscles Flexor Carpi Ulnaris, Palmaris Longus, and Flexor Capri Radiallis (6, 7, and 8, respectivelly). Thus, with three channels and three features (MAV, RMS, and WL) per channel, the system with nine inputs and one output is represented by the MISO model shown in Fig. \ref{miso model}, where its structure is composed by the schematic diagram of the deterministic parameters of the state-space model presented in (\ref{Equation2}) and (\ref{Equation3}), and by the schematic diagram of the Kalman filter presented in (\ref{Equation4_18}) and (\ref{Equation4_19}).
	
	The order of the model was defined by the residual variance error combination and the model order, with the Akaike Information Criterion (AIC) method \cite{haber1990structure}, as follows: 
	
	\begin{equation} \label{Equation4_14}
		{AIC} = \left( {1 + \frac{{2p}}{N}} \right){\sum\limits_{k = 1}^N {[y(k) - \hat y]} ^2},
	\end{equation}
	where $N$ is the number of observations and $ p $ is the number of parameters estimated by the model \cite{haber1990structure}.
	
	The AIC indicates the best order among the estimated models based on the lowest estimation residue associated with the goodness-of-fit of the estimated parameters data \cite{haber1990structure,ljung1996development}. Based on AIC for the order $n$ assessed from 1 to 12, verify that from the fourth-order onwards ($ n \geq 4 $), there was no significant decrease in the AIC index.
	
	We used 50$\%$ of the data set in the identification procedure, leaving the other 50$\%$ for the validation stage.
	
	%\subsection{MLP and NARX models}
	\subsection{Realization of existing grasping force methods}
	This subsection presents the utilization approach of the MLP network, the NARX model, and the LDA algorithm with a quadratic polynomial fitting, present in the existing methods in the literature, such as in \cite{choi2010real}, \cite{zhuojun2015semg},\cite{kumar2010adaptive}, \cite{ohno2017motion}, \cite{zhang2017pattern} and \cite{wang2019recognition}, in order to assess the performance and compare to the proposed approach.

	% A quadratic polynomial fitting (QPF) 
	\subsubsection{MLP model}
	The MLP network is a feed-forward network with at least three layers: an input layer, a hidden layer, and an output layer \cite{haykin2009neural}. 
	
\begin{figure}[thpb]
		\centering
		\includegraphics[scale=0.42]{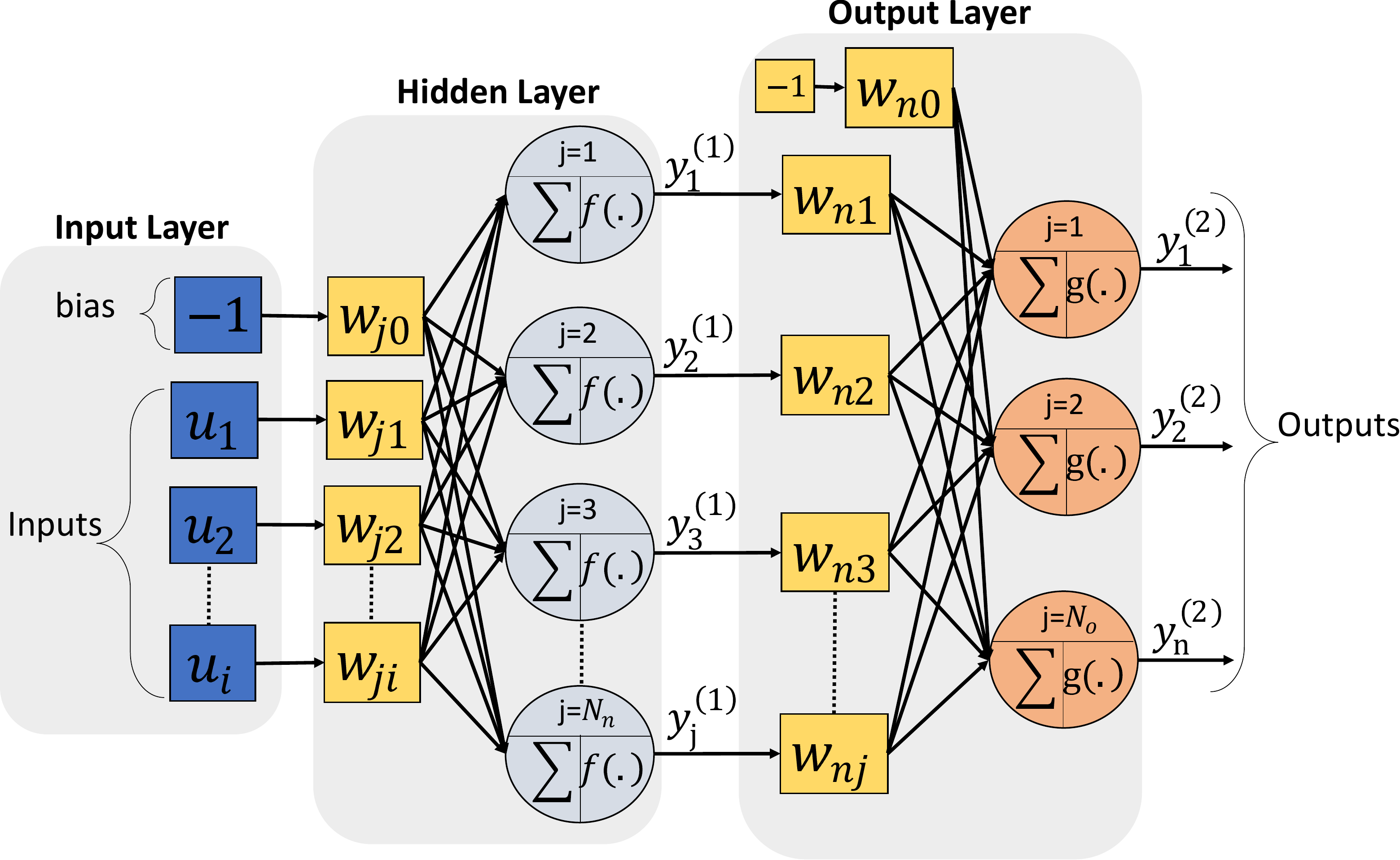}
		\caption{Structure of the MLP network.}
		\label{mlp_network}
	\end{figure}

	The MLP structure is shown in Fig. \ref{mlp_network} and the hidden and output layers are defined by:
	
	\begin{equation} \label{Equation NN_4_1}
		\begin{array}{l}
			y_j^{(1)} = f(\sum\limits_{i = 0}^{{N_i}} {{u_i}{w_{ji}}})\\
			j = 1,2,3 \ldots {N_n}
		\end{array}
	\end{equation} 
	\begin{equation} \label{Equation NN_4_2}
		\begin{array}{l}
			y_n^{(2)} = g(\sum\limits_{j = 0}^{{N_o}} {y_j^{(1)}{w_{nj}}} )\\
			n = 1,2,3 \ldots {N_o}\\
			
		\end{array}
	\end{equation}
	where $N_i$ is the number of neurons in the input layer, $bias$ is an activation threshold added to input 0, $u_i$ is the input vector, ${w_ {ji} ^ {(1)}}$ is the synaptic weight of the connections from the input layer to the hidden layer, $f (.)$ is the activation function of the hidden layer, $N_n$ is the number of neurons in the hidden layer, $y_j ^ {(2)}$ is the output vector of the hidden layer, $N_o$ is the number of neurons in the output layer, $ g (.) $ is the activation function of the output layer, ${w_ {nj} ^ {(2)}}$ is the synaptic weight of connections from the hidden layer to the output layer, and $y_n ^ {(2)}$ is the vector of outputs.
	
	The same input and output vectors used with the SS model were used to train the MLP network. To identify the grasping force model with the MLP network, we used three layers, and the sigmoid and linear activation functions, respectively, for the hidden and output layer. For the training of the neural network, we used the backpropagation algorithm optimized by the RMSprop method \cite{hinton2012neural}, with a maximum number of epochs equal to 1000 and mini-batch equal to 1. The mean square error (MSE) was set as a cost function, and the number of neurons in the hidden layer is equal to the order of the SS model ($ N_n= 4 $). 
	
	\subsubsection{NARX model}
	The discrete-time NARX model shows a good performance in modelling nonlinear systems and time series \cite{chen1990non,haykin2009neural}. This model can be mathematically represented as:  
	\begin{equation} \label{Equation_narx}
		y = f\left( {u(k),u(k - 1), \cdots, u(k - {n_u}),y(k - 1), \cdots, y(k - {n_y})} \right)
	\end{equation}
	where $y(k)$ and $u(k)$ represent the input and output at discrete time step $k$; $n_u$ and $n_y$ represent, respectively, the input- and the output-memory, and $f (.)$ is a nonlinear function.
	
\begin{figure}[!htbp]
		\centering
		\includegraphics[scale=0.35]{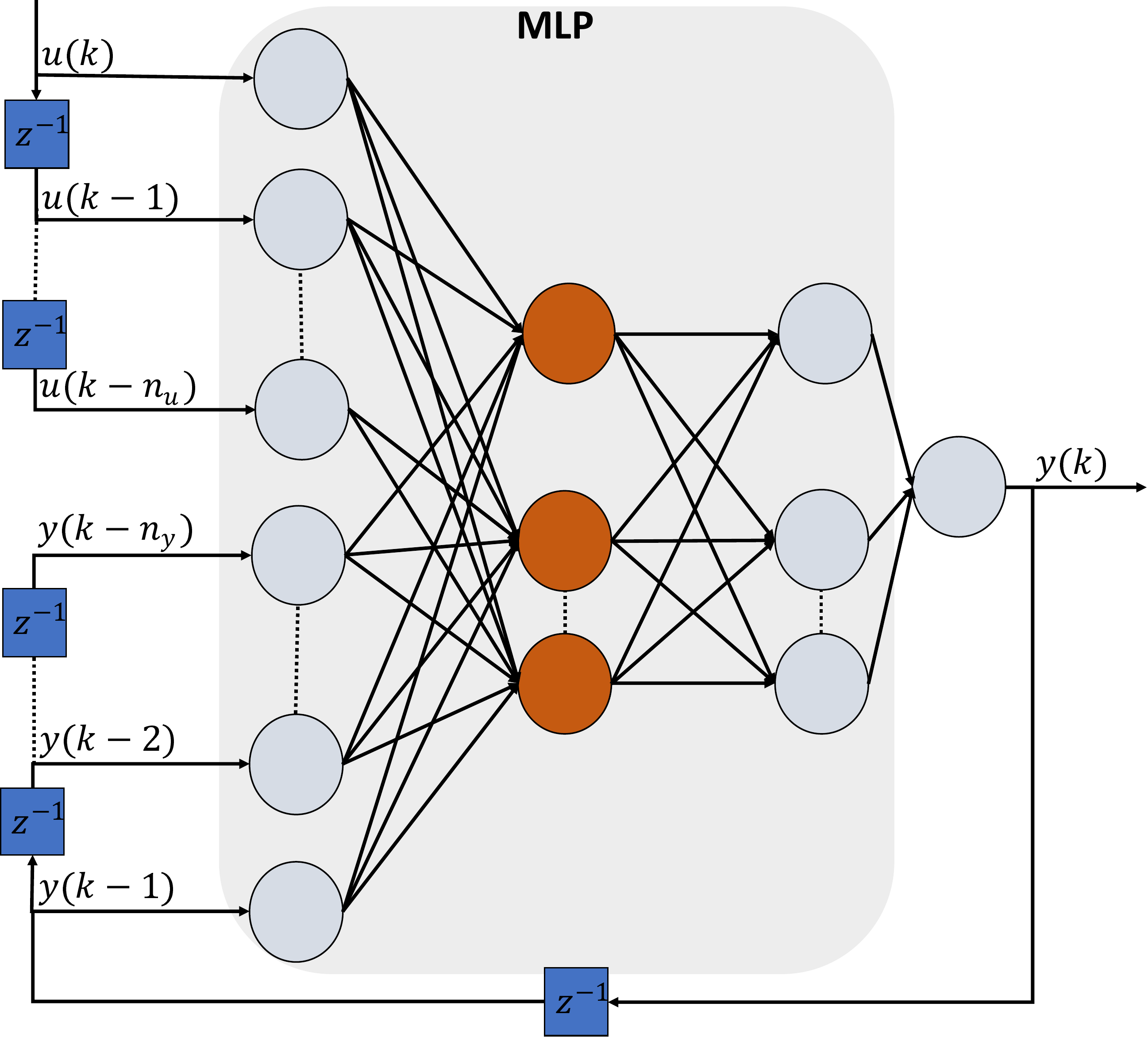}
		\caption{Structure of the NARX network.}
		\label{narx_network}
	\end{figure}

	In the present work, the NARX model is a recurrent dynamic neural network, which uses the structure of the MLP to find the nonlinear function $f$ that correlates the input $u$ and output $y$ of the model, as shown in Fig. \ref{narx_network}.
	To estimate the grasping force with the NARX-NN, we used the same parameters of the MLP network. Furthermore, the number of delays of the input signals and the delays of the model's output feedback signals were set equal to the order of the SS model.
	\subsubsection{LDA model}
	
	LDA is a well-known machine learning algorithm for classification and dimensionality reduction \cite{tang2005linear}. The basic idea of LDA is to find a projection matrix $W$ based on a linear combination of features, that transforms the original vector-matrix into a lower-dimensional space, which can separate and characterize two or more classes by simultaneously minimizing the within-class distance and maximizing the between-class distance \cite{hargrove2010multiple,wang2019recognition}. The mathematical formulation of the LDA algorithm was described by \cite{tang2005linear}. The dimensionality reduction  projection is represented by:

	\begin{equation} \label{Equation LDA}
		y = {W^T}x
	\end{equation}
	where the matrix y is the projected vector in lower-dimensional space and $x$ is the vector-matrix in the original dimension space.
	
	To estimate the grasping force, we used an approach where the LDA is used to reduce the dimension of the feature's input vector (MAV, RMS, and WL) to a one-dimensional vector-matrix. Subsequently, the QPF is used to find a polynomial model, which represents the relationship between the one-dimensional vector-matrix and the continuous grasping force, as proposed in \cite{wang2019recognition}.
	
	\section{Experimental results and discussion}
	\label{results}
	%In this section we present and discuss the results of continuous grasping force estimation with the proposed model in SS with KF. The model output is presented in Newton's force scale, and the model evaluation is made with statistical metrics to compare the estimated results and the measured outputs. It is also compared to the proposed regression model's result with other consolidated models, such as the model using MLP networks and the model using a neural network with the NARX structure.
	
	\subsection{Evaluation metrics}
	To evaluate estimation errors and the performance of the force estimation algorithms, the following indicators were used:  Normalized-Root-Mean-Squared-Error and the square of the Pearson’s correlation coefficient:
	\begin{equation} \label{Equation4_15_1}
		NRMSE=1-\frac{{\sqrt {\sum\limits_{k = 1}^N {{{\left[ {y(k) - \hat y(k)} \right]}^2}} } }}{{\sqrt {\sum\limits_{k = 1}^N {{{\left[ {y(k) - \bar y} \right]}^2}} } }},
	\end{equation}
	%and the coefficient of multiple correlation
	
	\begin{equation} \label{Equation4_15}
		{R^2} = 1 - \frac{{{{\sum\limits_{k = 1}^N {\left[ {y(k) - \hat y(k)} \right]} }^2}}}{{{{\sum\limits_{k = 1}^N {\left[ {y(k) - \bar y} \right]} }^2}}},
	\end{equation}
	where $ \bar y $ is the mean of $ N $ identification samples and $\hat y(k) $ is the estimated output.
	
	We considered values of $R^2$ above 0.80 to indicate a good representation of measured data by the model, with $R^2=1$ indicating its exact representation. For the NRMSE, the goodness of fit of the model has a maximum value of 1.
	
	\subsection{Continuous estimation and analysis of grasping force}
	
	%Figure \ref{r2_metric}.A shows that the parameters MAV, RMS, and WL have similar $R^2$ and NRMSE values $0.796 \pm 0.0611$ and $0.547\pm 0.0733$, respectively. Figure \ref{r2_metric}.B shows that the KF improves the estimation of continuous grasping force, considering both $R^2$ ($0.9066 \pm 0.044$ versus $0.9212 \pm 0.0319$) and NRMSE ($0.7 \pm 0.0718$ versus $0.7227 \pm 0.0563$).
	
\begin{figure}[!htbp]
		\centering
		\includegraphics[scale=0.65]{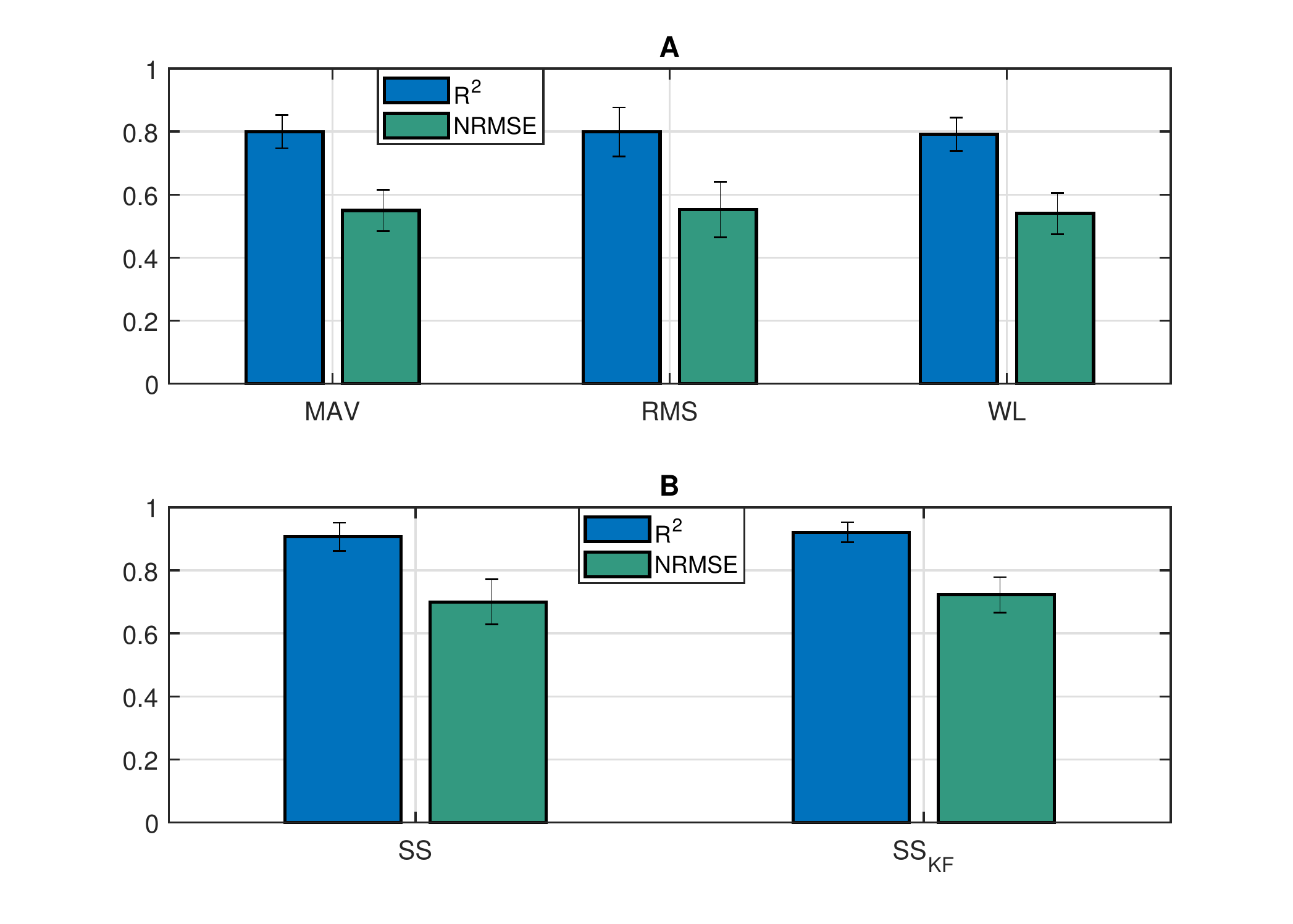}
		\caption{Average and standard deviation results of $R^2$ and NRMSE. (\textbf{A}) Results of the features MAV, RMS and WL. (\textbf{B}) Results of $R^2$ and NRMSE of the SS model with and without the Kalman Filter.}
		\label{r2_metric}
	\end{figure}

	The Fig. \ref{r2_metric}.A shows that the parameters MAV, RMS, and WL have similar results for $R^2$ and NRMSE metrics, with a mean of $0.796 \pm 0.0611$ to the $R^2$ and $0.547\pm 0.0733$ to the NRMSE. Thus, it can be seen in Fig. \ref{r2_metric}.B, that the SS model, in comparison with the direct usage of the presented parameters, shows an improvement in the multiple correlation index $(R^2=0.9066 \pm 0.044 )$ and in the normalized-root-mean-squared-error $(NRMSE=0.7 \pm 0.0718)$. Furthermore, the KF corrects the estimation errors and increases the performance of the SS model on estimating the continuous grasping force, improving the average and the standard deviation (SD) of the metrics results $(R^2=0.9212 \pm 0.0319, \, NRMSE= 0.7227 \pm 0.0563)$.
	
\begin{figure}[!htbp]
		\centering
		\includegraphics[scale=1.1]{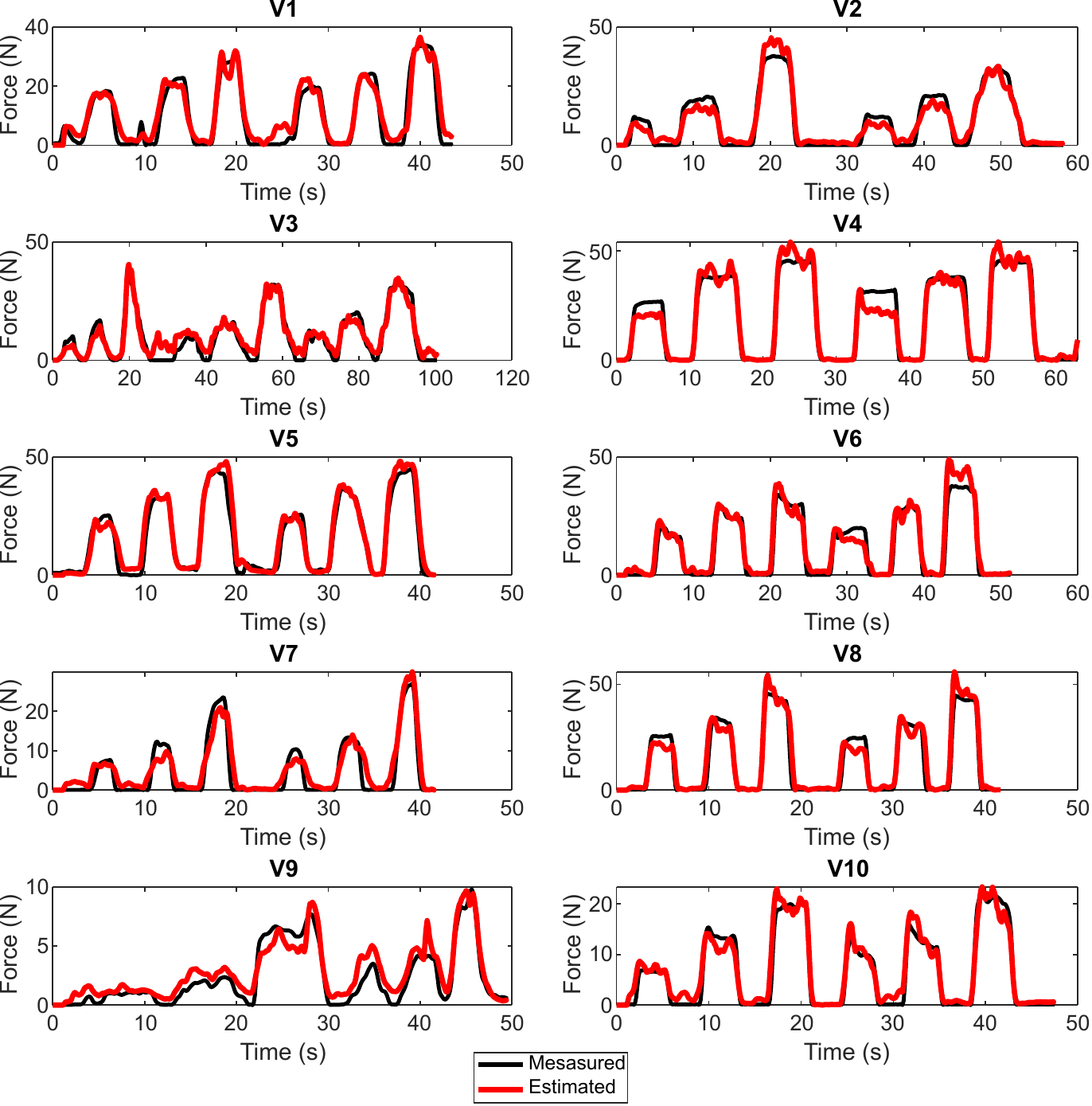}
		\caption{Grasping force estimation results with the SS$_{KF}$, for each voluntary.}
		\label{regression}
	\end{figure}

	From Fig. \ref{regression} it is possible to observe that the proposed model has a good performance on estimating the grasping force from sEMG features.  
	
	\subsection{Performance comparison of the SS model with other methods}
\begin{figure} % htb
		\centering
		\includegraphics[scale=0.62]{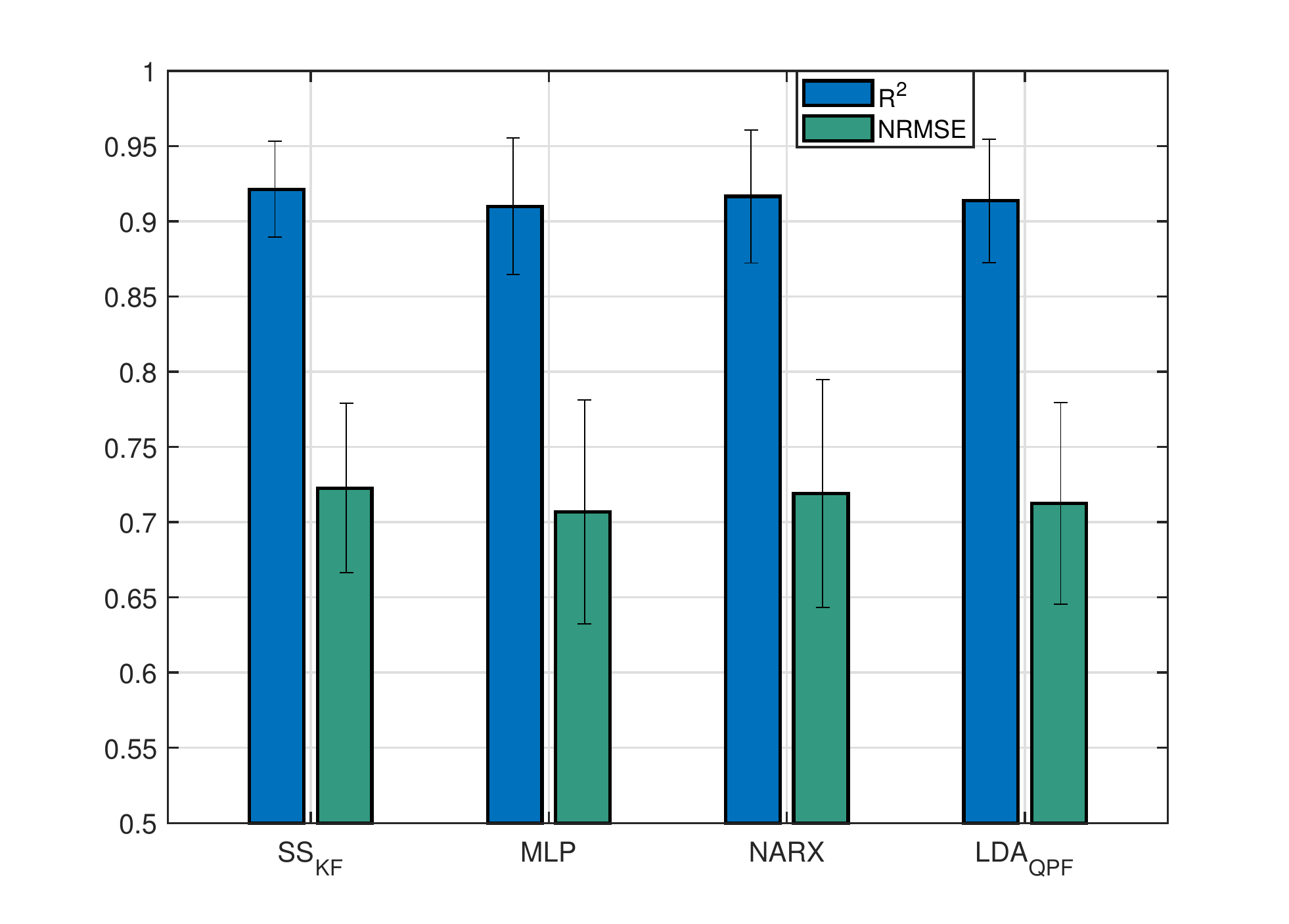}
		\caption{Results of $R^2$ and NRMSE to grasping force estimation from the models SS$_{KF}$, MLP, NARX and LDA$_{QPF}$.}
		\label{comparison_R2_regressors}
	\end{figure}

	We compared the performance of the SS model with the MLP, NARX and LDA$_{QPF}$ algorithms, which are commonly used in similar studies present in literature \cite{choi2010real,zhuojun2015semg, kumar2010adaptive, ohno2017motion, zhang2017pattern, wang2019recognition}. The results shown in Fig. \ref{comparison_R2_regressors} and Table \ref{table_acc_f1_class_indiv}, indicate that the models MLP, NARX, LDA$_{QPF}$, and the SS$_{KF}$ had similar results, with $R^2$ above $0.9$ and NRMSE above $0.7$. However, the SS$_{KF}$ had the highest $R^2$ and NRMSE. Moreover, both the $R^2$ and NRMSE of the SS$_{KF}$ model had the lowest variance, confirming the fact that the SS$_{KF}$ model was more precise and provided the lowest estimation error in our experiments.
	
\begin{figure} % htb
		\centering
		\includegraphics[scale=0.8]{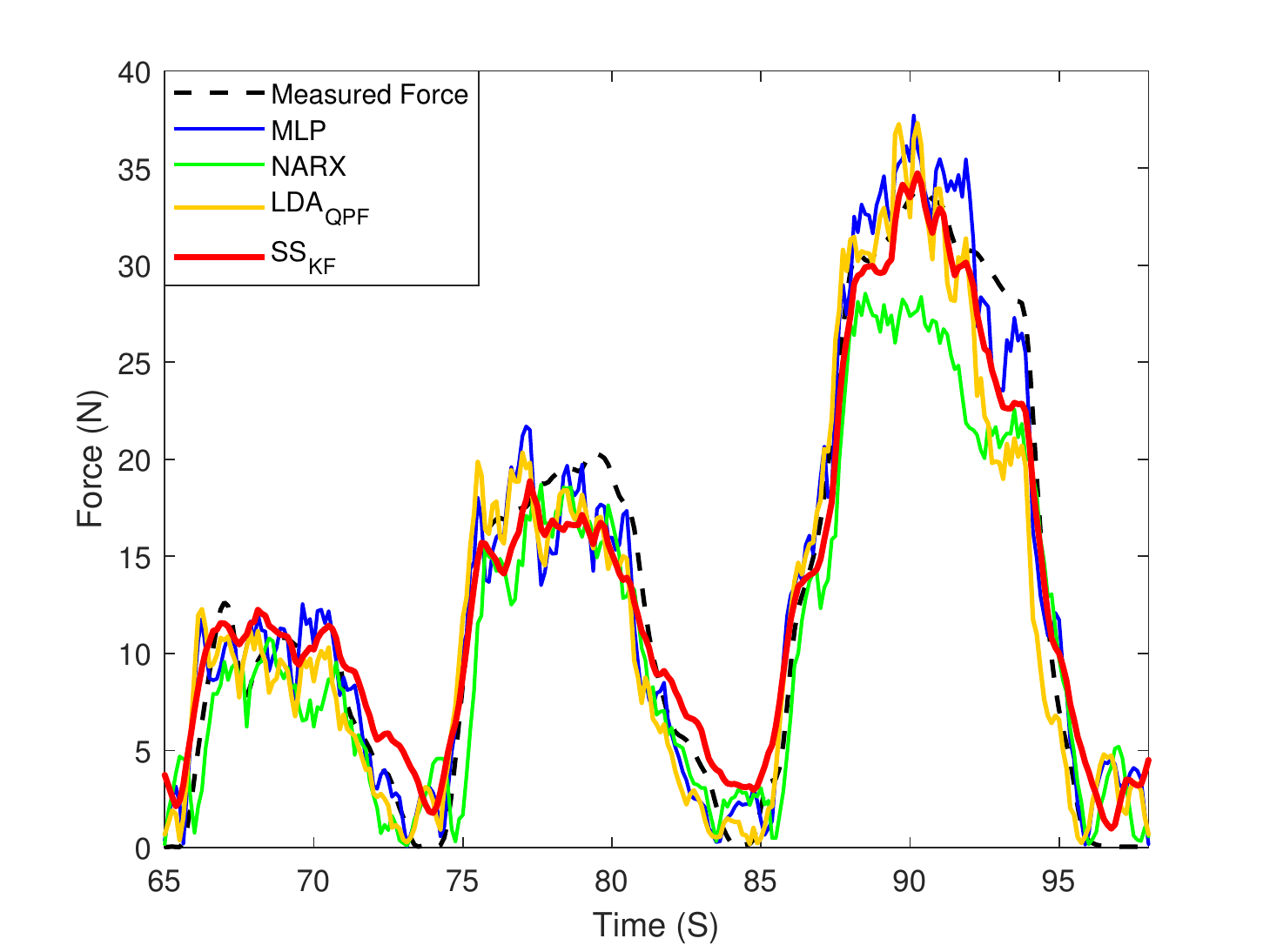}
		\caption{Grasping force estimation of voluntary 3, using the models MLP, NARX, LDA$_{QPF}$ and S$_{KF}$.}
		\label{comparison_ind2_regressors}
	\end{figure}

	The presented results indicate that the proposed SS$_{KF}$ model had the best performance and accuracy when estimating grasping force due to its low estimation error, low volatility and high precision. Despite all four models could be accepted as good choices for this estimation problem, the Figure \ref{comparison_ind2_regressors} and Table \ref{table_acc_f1_class_indiv} indicates that the SS$_{KF}$ model has a better goodness of fit compared to the LDA$_{QPF}$, NARX and MLP models.
	
	Table \ref{times_table} shows that the SS$_{KF}$ model has the fastest training time and both SS$_{KF}$ and $LDA_{QPF}$ stood out for having the lowest loop execution times, with  a loop runtime average below 0.01 ms, for working with real-time applications, proving to be models of low computational complexity. Also, our proposed model is advantageous because it is best suited for analysis based on control systems theory. Thus, with the possibility to assess stability issues by means of the identified models and their pole locations into the unitary circle, proving the technique to be based on stable, controllable and observable systems.
	
	\begin{table}
		\resizebox{\textwidth}{!}{\begin{tabular}{rlrrrrrrr}
				\hline
				Voluntaries &
				\multicolumn{2}{c}{SS\_KF} &
				\multicolumn{2}{c}{MLP} &
				\multicolumn{2}{c}{NARX} &
				\multicolumn{2}{c}{LDA} \\ \cline{2-9} 
				\multicolumn{1}{l}{} &
				\multicolumn{1}{c}{R²} &
				\multicolumn{1}{c}{NRMSE} &
				\multicolumn{1}{c}{R²} &
				NRMSE &
				\multicolumn{1}{c}{R²} &
				NRMSE &
				\multicolumn{1}{c}{R²} &
				NRMSE \\ \hline
				V1      & 0.90617 & 0.6919   & 0.84759 & 0.60641 & 0.83113 & 0.58778 & 0.90513  & 0.69198 \\ \hline
				V2      & 0.91252 & 0.70311  & 0.87473 & 0.64598 & 0.87539 & 0.64521 & 0.88021  & 0.65197 \\ \hline
				V3      & 0.94191 & 0.75803  & 0.94277 & 0.76073 & 0.95003 & 0.77618 & 0.94686  & 0.76948 \\ \hline
				V4      & 0.8761  & 0.64634  & 0.885   & 0.66072 & 0.88386 & 0.65828 & 0.88641  & 0.66273 \\ \hline
				V5      & 0.86109 & 0.61933  & 0.83272 & 0.58468 & 0.8918  & 0.6686  & 0.82408  & 0.58055 \\ \hline
				V6      & 0.93978 & 0.7546   & 0.93401 & 0.7424  & 0.94219 & 0.75911 & 0.9639   & 0.80998 \\ \hline
				V7      & 0.95433 & 0.78678  & 0.952   & 0.78091 & 0.96634 & 0.81624 & 0.94259  & 0.76034 \\ \hline
				V8      & 0.93114 & 0.7368   & 0.93758 & 0.75016 & 0.95473 & 0.78723 & 0.92541  & 0.72669 \\ \hline
				V9      & 0.93607 & 0.74814  & 0.94734 & 0.77033 & 0.91806 & 0.71305 & 0.9337   & 0.74251 \\ \hline
				V10     & 0.95258 & 0.78181  & 0.9449  & 0.76526 & 0.9517  & 0.77961 & 0.92708  & 0.72984 \\ \hline
				Average & 0.92126 & 0.72268  & 0.90986 & 0.70676 & 0.91652 & 0.71913 & 0.91354  & 0.71261 \\ \hline
				SD      & 0.0319  & 0.056333 & 0.04537 & 0.07440 & 0.04425 & 0.07571 & 0.041017 & 0.06701 \\ \hline
		\end{tabular}}

		\caption{Metrics result of the $ R^2 $ and NRMSE for each voluntary, obtained from the models  SS$_{KF}$, MLP, NARX, and LDA$_{QPF}$.}
		\label{table_acc_f1_class_indiv}
	\end{table}

	\begin{table}
		\resizebox{\textwidth}{!}{\begin{tabular}{lccccl}
				Model & \multicolumn{1}{l}{Training time (ms)} & \multicolumn{1}{l}{Loop Runtime (ms)} & \multicolumn{1}{l}{Controllable} & \multicolumn{1}{l}{Observable} & Stable \\ \hline
				SS\_KF   & 6.97 $\pm$ 3.11                       & 0.00735 $\pm$  0.0035                   & yes & yes & yes                   \\ \hline
				NARX     & 26.5 $\pm$  21.9                      & 0.0338 $\pm$ 0.014                     & -   & -   & \multicolumn{1}{c}{-} \\ \hline
				MLP      & 1572 $\pm$  2106                      & 0.010 $\pm$   0.0033                   & -   & -   & \multicolumn{1}{c}{-} \\ \hline
				LDA\_QPF &  15.37 $\pm$ 10.6 &  0.00742$\pm$  0.0032 & -   & -   & yes                  
		\end{tabular}}
		
		\caption{Results of training time, loop runtime, and the availability of assessing the properties of controllability, observability and stability of the models SS$_{KF}$, MLP, NARX, and LDA$_{QPF}$ }
		\label{times_table}
	\end{table}

	\subsection{Real-time implementation}
	
	For implementation in prosthetic systems, the proposed model can be embedded in mini computer systems such as the raspberry PI and microcontrollers, with minimum requirements for computational power. The estimated grasping force can also be visualized in real-time for use in rehabilitation and diagnostic systems. In this section, we show the results of proof-of-concept experiments.

\begin{figure}
		\centering
		\includegraphics[scale=0.85]{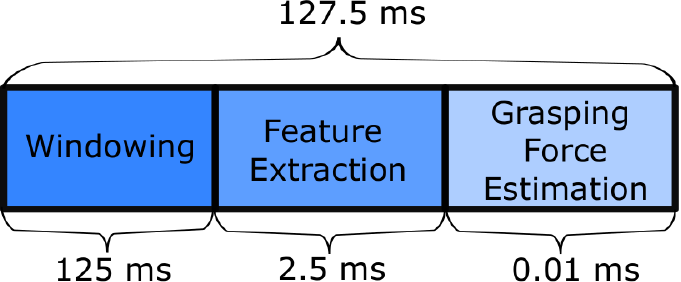}
		\caption{Processing time of all techniques used in the grasping force system estimation.}
		\label{times_}
	\end{figure}

	In the experimental tests, a total processing time of 127.5 ms was observed for windowing, feature extraction, and grasping force estimation, as shown in Fig. \ref{times_}. In a time interval of 300 ms, the system can provide up to two grasping force estimations. Thereby, a prosthetic device based on the proposed model can allow the user to handle objects in daily life activities without noticeable operational delay. 
	
\begin{figure}
		\centering
		\includegraphics[scale=0.45]{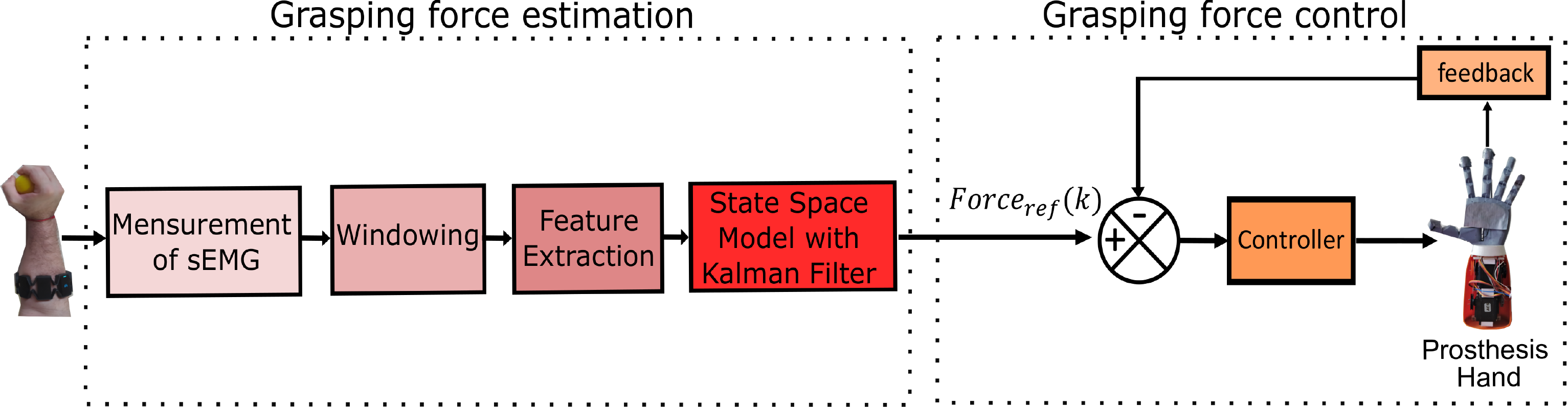}
		\caption{Prosthesis' grasping force control architecture.}
		\label{architecture}
	\end{figure}

	For grasping force control of prosthetic systems, the system can be implemented according to the architecture presented in Fig. \ref{architecture}. 
	In applications that require the direct use of estimated continuous and proportional grasping force, the reference signal should not contain a spurious shape and a high level of volatility, since these adversely affect the gain margin and the phase margin of the control system. Thereby, the SS model with Kalman Filter proves to be more advantageous because it presents an output with minimum variance and lower volatility, as shown in Fig. 11 and Table 2, avoiding the use of post-processing filters that can cause processing delays.
	
	\section{Conclusion}
	In the present work, we proposed a state-space model with a Kalman Filter to continuously estimate grasping force and that could be used as input to control the movement of a robotic hand during manipulation tasks. The performance of the system was evaluated with data recorded from ten experimental subjects. The sEMG signals of the subjects were recorded while they manipulated an object with three different grasping force magnitudes. We used the RLS algorithm for the identification of the state-space model. The SS model was able to estimate the grasping force from an input array composed of the features MAV, RMS, and WL from sEMG signals recorded from three forearm's flexor muscles. Moreover, the use of a Kalman Filter tuned with the minimum variance case improved the performance of the SS model.
	
	In proof-of-concept experiments, the SS model with KF had satisfactory performance results with a low computational complexity in real-time applications. Thus, based on several performance indexes, the presented results indicate that the proposed model is a better option for the real-time estimation of grasping force from sMEG recordings than some regression methods, such as the MLP network, NARX model, and the LDA algorithm with quadratic polynomial fitting.
	
	The proposed system proved to be both stable and accurate for real-time applications and feasible for embedding in micro-controlled systems to work in myoelectric prosthesis.
	
	\label{conclusion}
	
	
	
	%% References
	
	\section*{Acknowledgment}
	The authors thankfully acknowledge the financial support of the Brazilian National Council for Scientific and Technological Development (CNPq) under grant 142415/2018-9.
	
	\bibliographystyle{elsarticle-num}
	\bibliography{mainbib}
	
	
	
	%\begin{thebibliography}{00}
	
	
	
	%\end{thebibliography}

\end{document}